\documentclass[11pt]{article}

\usepackage{amsmath, amssymb}
\usepackage[mathscr]{eucal}

\usepackage{graphicx}          
\usepackage{subfigure}

\usepackage[usenames,dvipsnames]{color}

\allowdisplaybreaks
\newtheorem{theorem}{Theorem}

\newtheorem{proposition}{Proposition}
\newtheorem{corolary}{Corollary}

\usepackage[paper=a4paper,
            lmargin=3.0cm,rmargin=2.9cm,tmargin=3.8cm,bmargin=3.0cm,%
            foot=0.9truecm]{geometry}
\begin{document}

\baselineskip14pt
\title{A new look at the Feynman `hodograph' approach   to the Kepler first law. }

\author{ 
Jos\'e F.\ Cari\~nena$\dagger\,^{a)}$,
Manuel F.\ Ra\~nada$\dagger\,^{b)}$, and
Mariano Santander$\ddagger\,^{c)}$ \\ [2pt]
$\dagger$
   {\sl Departamento de F\'{\i}sica Te\'orica and IUMA, Facultad de Ciencias} \\
   {\sl Universidad de Zaragoza, 50009 Zaragoza, Spain}  \\   [2pt]
$\ddagger$
   {\sl Departamento de F\'{\i}sica Te\'{o}rica and IMUVa, Facultad de Ciencias} \\
   {\sl Universidad de Valladolid, 47011 Valladolid, Spain} 
} 
\date{January 7, 2016}
\maketitle 

 
\begin{abstract}
Hodographs for the Kepler problem are circles. 
This fact, known since almost two centuries ago, still provides the simplest path to derive the Kepler first law. 
Through Feynman `lost lecture', this derivation has now reached to a wider audience. 
Here we look again at Feynman's approach to this problem as well as at the recently suggested modification 
by van Haandel and Heckman (vHH), with two aims in view, both of which extend the scope of the approach. 

First we review the geometric constructions of the Feynman and  vHH approaches 
(that prove the existence of {\itshape elliptic} orbits  without making use of integral calculus or differential equations) 
and then we extend the geometric approach to cover also the {\itshape hyperbolic} orbits 
(corresponding to $E>0$). 
In the second part we analyse the properties of the director circles of the conics, which are used to simplify the approach and we relate with the properties of the hodographs and with  
the Laplace-Runge-Lenz vector, the constant of motion specific to the Kepler problem. 
Finally, we briefly discuss the generalisation of the geometric method to the  Kepler problem in configuration 
spaces of constant curvature, i.e. in the sphere and the hyperbolic plane.

\end{abstract}
	
\begin{quote}
  Geometry of the Kepler problem.  Kepler laws.  Hodographs.   
Laplace-Runge-Lenz vector.  Conics. Director circle of a conic.  


\end{quote}

\vfill
\footnoterule
{\noindent\small
$^{a)}${\it E-mail address:} {jfc@unizar.es } \\
$^{b)}${\it E-mail address:} {mfran@unizar.es }  \\
$^{c)}${\it E-mail address:} {msn@fta.uva.es }
}
\newpage

\section{Introduction}

The Kepler problem, i.e., the motion of a particle under a inverse square law, has been a true landmark in physics. Since antiquity the general assumption was that planets moved in circles, an idea shared by Copernicus himself.  
However, Kepler, analysing a long series of astronomical observations found a very small anomaly in the motion of Mars, and that was the starting point for his discovery that the planet orbits were ellipses.  However, there is some historical irony in the fact that a circle is still the exact solution of a related view of the problem, when any Kepler motion is seen not in the ordinary configuration space, but in the `velocity space'. 

When a point particle moves, its velocity vector, which is tangent to the orbit, changes in direction as well as in modulus. 
We might imagine this vector translated in the naive manner to a fixed point. Then, as the particle moves along its orbit, the tip of the velocity vector traces a curve in velocity space that Hamilton called `hodograph' of the motion, to be denoted here by ${\mathcal H}$. In Hamilton's own words \cite{Ham1847}: 

\begin{quotation}
\noindent{\sl\footnotesize \dots the curve which is the locus of the ends of the straight
lines so drawn may be called the hodograph of the body, or of its
motion, by a combination of the two Greek words, $o\delta o\zeta$,
a way, and $\gamma\rho\alpha\phi\omega$, to write or describe;
because the vector of this hodograph, which may also be said to be
the vector of velocity of the body, and which is always parallel
to the tangent at the corresponding point of the orbit, marks out
or indicates at once the direction of the momentary path or way in
which the body is moving, and the rapidity with which the body, at
that moment, is moving in that path or way. }
\end{quotation}

The statement of the circularity of Kepler hodographs is an outstanding example of the rediscovery of a wheel (as pointed out in \cite{Derb01}); its first statement can be traced back to the 1840's independently to M\"obius \cite{Mob1843} and Hamilton \cite{Ham1847}, to be later rediscovered several times by many authors including Feynman. By putting this property at the outset one can obtain a complete solution for the shape of the orbits with a minimum of additional work. Thus, the common idea to deal with the Kepler motion in all these `indirect' approaches is to start by a proof of the circular character of hodographs and afterwards to derive the conic nature of Kepler orbits. 

For an  historical view of this question we refer to a paper by Derbes \cite{Derb01} which also gives a very complete discussion of the problem in the language of classical Euclidean geometry, including the contribution to this very problem of outstanding figures as J.C. Maxwell \cite{Max52}. The  historical constructions are extended in this paper even to parabolic orbits (see also the paper \cite{Kow03}). 

The hodograph circular character for the Kepler problem is closely related to the existence of an {\itshape specifically Keplerian} constant of motion which which is an {\itshape exceptional property of the central potential with radial dependence $1/r$}. From a purely historic viewpoint, this vector can be traced back to the beginning of XVIII century, with J. Hermann and J. Bernoulli (see two notes by Goldstein  \cite{Gold75, Gold76}), being later rediscovered independently several times. The connection  with the circular character of the hodograph seems to be due to Hamilton \cite{Ham1847}; from a modern viewpoint all these distinguished  properties are linked to the superintegrability of the Kepler problem (for a moderately advanced discussion, see \cite{CRS08JPA}).

In a recent paper \cite{vHH09}, van Haandel and Heckman (hereafter vHH) have pushed this `Feynman's construction' a further step, providing a fully elementary proof of the elliptic nature of the (bounded) Kepler orbits. In the form presented by vHH, this applies only for {\itshape non degenerate} (angular momentum $L\neq 0$) {\itshape elliptical} orbits (with $E<0$, and thus bounded). In this paper we first prove that a quite similar construction is also valid for the unbounded $E>0$ hyperbolic orbits. This requires some restatement of the vHH results, along which some circles, the {\it director} circles of the Kepler orbit as a conic, appear. When the role of these circles is properly recognised, the vHH derivation can be streamlined and presented in a way more clear than the original and simpler than the Feynman one.

This is the plan of this paper: A short introductory section serves to state the problem and to set notation so as to make the paper self-contained. A brief description of both the Feynman and the vHH approaches for elliptic orbits foolows; a particular Euclidean circle underlies both approaches.  Then we discuss a reformulation of the vHH approach, where the basic properties of this Euclidean circle are as clearly  stated  as possible. Once the real geometric role played by this circle has been identified, the extension to hyperbolic orbits can be performed easily  (we refer to \cite{RanSan08} for some complementary details). This new, slightly different, construction is streamlined in the next Section. The `reverse part', which goes from the hodograph to the configuration space orbit is also fully characterised and studied; it turns out to be a  bit simpler than Feynman and  vHH construction. 

All this will cover only the Euclidean Kepler problem. In the last section we briefly indicate how the `Kepler' problem in constant curvature spaces, i.e., on the sphere and on the hyperbolic plane, can be approached and solved following precisely the pattern described case in the previous section. The essential point in this connection is to deal with the momenta, instead of dealing with the velocities. Neither the Feynman one nor the vHH approach seem to allow such a direct extension.

\section{Problem statement and some notations}

The motion of a particle of mass $m$ in Euclidean space under a general 
conservative force field derived from a potential 
${\bf F}({\bf r})=-\bf\nabla V({\bf r})$ has the total energy 
$E$ as a constant of motion.  Units for mass will be chosen so that
$m=1$; after this choice the momentum ${\bf p}$ can be 
assimilated to the velocity vector ${\bf v}={\dot{\bf r}}$. 

When the force field is central (from a centre $O$),  angular momentum 
${\bf L}= {\bf r} \times {\bf p}$ is also conserved so the orbit is contained 
in a plane through $O$ (perpendicular to ${\bf L}$) and, if Cartesian coordinates 
are chosen so that  ${\bf L}=(0,0,L)$, then the motion is restricted to the plane $z=0$. 
From the point of view of this plane, $L$ appears as an scalar, which may
be either positive or negative. Constancy of $L$ is related to the
{\itshape law of areas} $r^2\dot{\phi}=L$ and leads to the 
second Kepler law, which holds for motion under {\it any central potential}.

The Kepler problem refers to the motion in Euclidean space of a
particle of mass $m$ under the central force field 
\begin{equation}
{\bf F}({\bf r})=-\frac{k}{r^2} \left(\frac{{\bf r}}{r}\right)
\end{equation}
(centre placed at the origin $O$), or equivalently, under the potential $V({\bf r})=-k/r$,  $k>0$. The main results for this problem are embodied in the Kepler laws, whose first
mathematical derivation was done by Newton in the Principia
\cite{NewPrin1687} (see also \cite{Chandra97}). The first law was
stated by Kepler as {\it the planet's orbits are ellipses with a
focus at the centre of force}. 
Actually not only ellipses, but also parabolas and one of the  branches  
of a hyperbola (with a focus at the origin) may also appear as orbits for an  
{\it attractive central force with a $1/r^2$ dependence}, 
and the general Kepler first law can be restated as saying that the {\itshape Kepler orbits are conics with a focus at the origin}.

The constructions to be discussed in this paper are made within
synthetical geometry, and we freely use the usual conventions: in Euclidean plane points are denoted by capital letters $O,P$ and symbols as $OP$ will denote either the line through
points $O$ and $P$ or the segment $OP$ seen as an (affine) vector, i.e.,
a vector at $O$ whose tip is at $P$: the modulus $|OP|$ of this
vector is the Euclidean distance between points $O$ and $P$ (see also
\cite{Derb01}). 

\subsection{The focus/directrix characterisation of Euclidean conics}

There are three types of (non-degenerate) conics in the Euclidean plane: two generic types, {\itshape ellipses} and {\itshape hyperbolas} and one non-generic type, {\itshape parabolas}. The two generic types, i.e., ellipses (resp.\  hyperbolas) are  geometrically characterised by the property 
\begin{quote}
{\itshape The sum (resp. the difference) of the distances from any point on the curve to two fixed points, called {\itshape foci}, is a constant}; 
\end{quote}
this property is behind the well known `gardener' construction of ellipses. For parabolas, one of these foci goes to infinity, so the previous characterisation degenerates, and must be replaced by another property, as, for example 
\begin{quote}
{\itshape The distances from any point on the parabola to a fixed line $D$ called {\itshape directrix} line and to a fixed point $O$, called focus, are equal}. 
\end{quote}
This characterising property can also be generalised to include ellipses and hyperbolas, as we will see next.

It turns out that the two foci of conics appearing in the Kepler problem plays different roles, and from the start we adapt our notation to this asymmetry: the two foci of the ellipses and hyperbolas will be denoted $O$ and  $I$, and the single focus of parabolas as $O$.  Ellipses and hyperbolas degenerate to parabolas when the {\itshape second focus} $I$ goes to infinity. 

An interesting but less known alternative characterisation also exists for ellipses and hyperbolas, which is based on a pair {\itshape focus-directrix}. For these two generic types of conics the directrix is not a straight line, but a {\itshape circle} called {\itshape director circle}. Thus ellipses (resp.\  hyperbolas)  can be characterised geometrically by the property \begin{quote}
{\itshape The distances from any point on the ellipse (resp.\  hyperbola) to a fixed circle, $\mathcal{D}_O$ and to a fixed point $O$ are equal}. 
\end{quote}
The two generic types of conics corresponds to the relative position of $\mathcal{D}_O$ and $O$: for an ellipse (resp.\  a hyperbola) the point $O$ is inside (resp.\  outside) 
the circle $\mathcal{D}_O$. 

There is not a fully standard naming for several circles associated to a conic, and therefore some confusion may follow. We stick here to the naming used by Sommerville, \cite{SomConics24} where {\itshape director circle} applies (for ellipses and hyperbolas) to a circle with centre at a focus, radius $2a$ and with the property that the points on the conic are equidistant from the other focus and from the director circle $\mathcal{D}$. 

Another circle is the {\itshape orthoptic circle}\cite{Pedoe}, which is defined as the set of points where two 
{\itshape perpendicular} tangents to the conic meet; it is easy to prove that for ellipses and for hyperbolas this set of points is also a circle. The name orthoptic refers to the fact that, when viewed from any point on this circle, the ellipse spans visually the interior of a right angle and the hyperbola spans part  of the exterior of a right angle. For parabolas, the set of points with this property degenerates to a straight line and turns out to coincide with the directrix, which partly explains why this circle is sometimes called  the director circle; as indicated before we are not following this usage. 

Ellipses and hyperbolas have two foci, and therefore two director circles, denoted $\mathcal{D}_I$ (resp. $\mathcal{D}_O$) which refer respectively to the circle with centre at the focus $O$ (resp. $I$(, radius $2a$ and with the property that the points on the conic are equidistant from the focus $O$ (resp. $I$) and from the corresponding director circle $\mathcal{D}_O$ (resp. $\mathcal{D}_I$). 

\begin{figure}[h]
\includegraphics[width=\hsize]{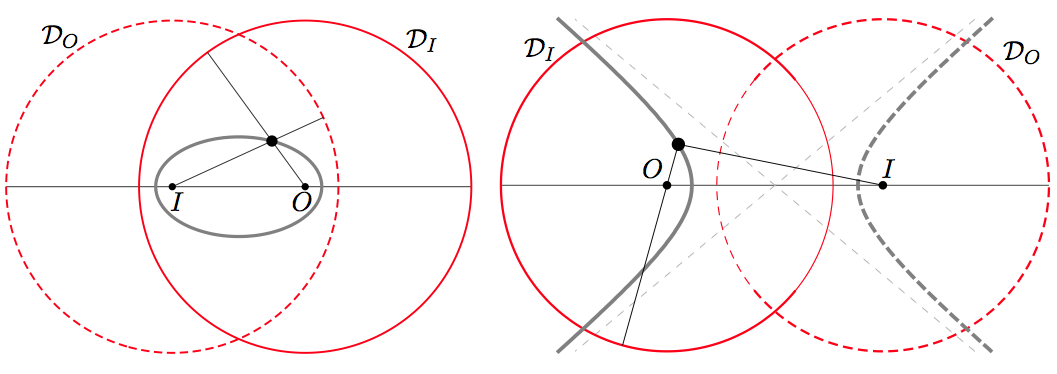}
\caption{\footnotesize \leftskip=20pt \rightskip=20pt 
Director circles for ellipses and hyperbolas. }
\label{FigDirectCircleHyperb}
\end{figure}

The equivalence of the `gardener' characterisation and the one based in the focus-director circle pair is clear. For ellipses and 
hyperbolas the two director circles $\mathcal{D}_O$ and $\mathcal{D}_I$ have its centers at the `other' focus $I$, $O$, and radius equal to the major axis $2a$.  A central symmetry in the ellipse or hyperbola centre swaps the two focus and the two director circles. The non-generic type of conics, parabolas, have no  centre and the radii of the two director circles, that must be equal, are infinite. In this case, (i) the focus $I$ goes to infinity, and with it the director circle $\mathcal{D}_I$ (with centre at $O$) goes also to infinity; and (ii)   the director circle $\mathcal{D}_O$ has centre at infinity, and appears as a straight line, which is the parabola directrix $D$.

Another basic property of Euclidean circles should be mentioned. Let a circle $\mathcal{C}$ and a fixed point $P$ be given in the plane. Consider all straight lines in that plane through $P$.

\begin{itemize}
\item   If $P$ is interior to $\mathcal{C}$ all these lines will intersect $\mathcal{C}$ in two (real) points.
 
\item    If the point $P$ is exterior to $\mathcal{C}$, then there will be two real tangent straight lines to $\mathcal{C}$ through $P$, and all straight lines within
 the wedge limited by the two tangents will intersect $\mathcal{C}$ in two real points. 
  \end{itemize}
  
In all these cases, if $d_1$ and $d_2$ denote the (oriented) distances from $P$ to the two intersection points along a particular straight line, then the product $d_1 d_2$   turns out to be independent of the chosen straight line. This value is called the {\itshape `power of the point $P$ relative to the circle $\mathcal{C}$}. This power    is negative if the point $P$ is inside $\mathcal{C}$ (then $d_1, d_2$ have opposite orientations) and is positive if $P$ is outside $\mathcal{C}$. A proof is    easily constructed and we leave it to the reader. 

\subsection{Some non-standard 2d vector calculus }

In a 2D plane, there is a canonical way to associate to any vector ${\bf w}$ another vector denoted  ${}^*{}{\bf w}$ (to be understood as a single symbol). 
This possibility is specific for a 2D plane and does not happen in the 3D space, where the `similar' construction, the vector product, requires to start from {\itshape two vectors}. The vector ${}^*{}{\bf w}$ is defined to be the (unique) vector in this plane orthogonal to ${\bf w}$,  with the same modulus as ${\bf w}$ and such that the pair $({\bf w}, {}^*{}{\bf w})$ is positively oriented. The vector ${}^*{}{\bf w}$ is obtained from ${\bf w}$  by a rotation in the plane by an angle $+\pi/2$,  and component-wise,  ${}^*{}w_j=\epsilon_{ij}w^i$ (with sum in the repeated index $i$), i.e., if ${\bf w} = (w^1, w^2)$, then ${}^*{\bf w} = (-w^2, w^1)$.  We now state two properties which are easy to check: 

\begin{itemize}

\item[a)]  ${}^*{}({}^*{}{\bf w})=-{\bf w}$. 

\item[b)] If ${\bf L}$ is a vector
perpendicular to this plane, then the vector product ${\bf L}\times {\bf w}$ 
can be expressed in terms of the modulus $L$ of ${\bf L}$ and of ${}^*{}{\bf w}$, as ${\bf L}\times {\bf w} = {L \,}^*{\bf w}$.
\end{itemize}

In the natural identification $(x, y) \equiv x + {\rm i}\, y$ of the Euclidean plane ${\mathbb R}^2$ with ${\mathbb C}$, the operator ${\bf w} \to {}^*{\bf w}$ corresponds to multiplication by the complex unit ${\rm i}$.

\section{The geometric approach to the Kepler first law}

\subsection{Feynman approach for elliptic orbits}

In 1964, Feynman delivered a lecture on {\itshape `The motion of planets around the
sun'} which was not included in the published {\itshape`Lectures on Physics'}. Feynman's notes
for this lecture were eventually found, and then published and commented by Goodstein and Goodstein in 1996 \cite{FeynmanLostLecture}. In his peculiar style, 
Feynman gave an elementary derivation of  Kepler first law by focussing attention in the hodograph. Such derivation starts by unveiling (rediscovering) 
a  curious property: the Kepler hodographs have an exact circular character, but this circle is not centred in the origin of velocity space (see e.g.   \cite{AbelEtAl75,Stic78a,Stic78b,GonEtAl98,But00,Apos03,Mor05}).

The publication of the `Lost Lecture' has made this approach to the Kepler problem more widely known than before, although, as Counihan points out in  \cite{Coun07},  this geometric approach was probably more in line with the background of XIX century mathematical physicists than it is nowadays.   

This procedure of studying the Kepler motion reduces to a minimum the resort to calculus or to differential equations.  All the `hodograph first' approaches to solve Kepler problem (Feynman's included)  require to establish first the circular nature of the Kepler hodograph.  Some resort---more or less concealed--- to solving a differential equation is required here.  The standard way is to write the Newton laws for the motion ${\bf x}(t)$ in a central field of forces with an  $1/r^2$ radial dependence and look for the differential equation satisfied for the velocity $\dot{\bf x}(t)$  (see e.g. Milnor \cite{Mil83},  where one can find a careful discussion). 

Newton had to solve this problem  by a geometrical argument involving a kind of discretisation of the problem, considering positions at {\itshape equispaced times} $t, t+\Delta t, t+2\Delta t, \dots$, and, 
as it is well known, this leads to a complicated description. 

But since Hamilton we know that this non-linear problem can be transformed to a linear one if we change the time $t$ by the angle $\phi$ as the independent variable 
and we then enforce the law of areas. The function $\dot{\bf x}(\phi)$  which gives the velocity in terms of the angle $\phi$ satisfies a {\itshape linear equation} whose solutions are immediately seen to be circles in the velocity space. Feynman solved this step by making a kind of discretisation similar to the one by Newton, but  involving {\itshape equispaced angular positions} $\phi, \phi+\Delta \phi, \phi+2\Delta \phi, \dots$ on the orbit. This provides  some kind of discrete analogous of the linear equation   satisfied by $\dot{\bf x}(\phi)$, and leads in the limit $\Delta\phi\to 0$ to the circular character of the hodograph. Once this 
fact has been established, Kepler first law follows in a simple and purely algebraic way. 

Of course, it remains to describe the relation among the hodograph and the orbit. We need a construction which applied to the hodograph would allow us to recover the orbit. In the Feynman lecture, even if rather informally presented, this is accomplished through a sequence of  three transformations, whose essential part is to rotate the hodograph by $-\pi/2$ around the origin $O$. All the necessary details will be given in the following sections,   after dealing with another recent construction, due to van Haandel and Heckman.

\subsection{van Haandel--Heckman approach for elliptic orbits}

Van Haandel and Heckman  \cite{vHH09} introduced a modification in the Feynman approach which reverses the standard  `hodograph approach', and even avoids the need to draw on a differential equation, thus providing a good way to present the problem to beginners. 
They compare their  derivation with the one devised by Feynman  and put both into perspective against the original Newton derivation. This comparison makes sense   because all three derivations are framed in the language of synthetical Euclidean geometry.

The geometric construction they propose has many elements in common with the previous ones (Maxwell, Feynman, \dots) but they look at the problem from a different perspective  which leads much more directly to two essential insights in the problem: the conic nature of the orbits {\itshape and} the existence of an `exceptional' Keplerian constant of motion ${\bf I}$. It is worth to    emphasise that the derivation  is purely algebraic, and at no stage a resort to a differential equation should be done (in contrast to Feynman approach). 

The standard Laplace-Runge-Lenz (LRL) vector ${\bf A}$ is known to point from the force centre to the perihelion, along  the orbit major axis, with modulus $k\,e$; otherwise ${\bf A}$ lacks any geometrical interpretation. On the contrary, the constant vector ${\bf I}$ which follows from this approach is a a rescaling of the standard LRL vector ${\bf A}$ by a factor $1/E$, ${\bf I}={\bf A}/{E}$ and admits a nice and direct geometrical interpretation: both for elliptic and hyperbolic orbits it goes from the force centre, which is one focus of the orbit, to the  `second' or `empty'  focus (it degenerates to an  infinite modulus vector along the conic axis for parabolic orbits). 

We start by recalling the elementary proof of Kepler
first law  as proposed by Van Haandel and Heckman in \cite{vHH09}. 
Consider Kepler orbits with $L\neq 0$ and $E<0$ 
(we already know  they are Kepler ellipses, but assume at this point  that we do not know this).  

As a consequence of energy conservation,  motion in configuration space (or in the plane of motion) is confined to the interior of a circle ${\mathcal D}$, centred at the origin and with radius $k/(-E)$. 
Outside this circle the kinetical energy would be negative, and thus this exterior region is forbidden for classical motion.
This circle $\mathcal D$ plays an important role (but as we shall see later, this role is not exactly as the boundary of energetically allowed region, though this is the way van Haandel--Heckman presented the construction).

\begin{figure}[h]
\includegraphics[width=\hsize]{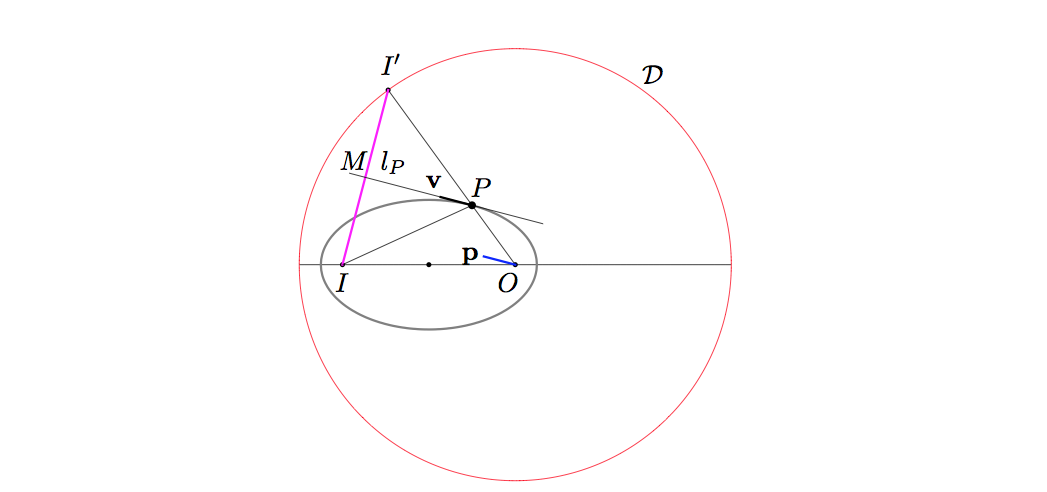}
\vskip-\baselineskip
\caption{\footnotesize \leftskip=20pt \rightskip=20pt 
The van Haandel-Heckman construction: $E<0$ Kepler orbits are ellipses}
\label{FigvHHelip}
\end{figure}

Let be ${\bf r}$  the  position vector of  a point $P$ on a given orbit,   $l_P$ denote the tangent line to the orbit at $P$, ${\bf v}$ the velocity of the particle at $P$
and ${\bf p}$ the linear momentum vector, which we will imagine as attached to the origin $O$, 
i.e., the vector ${\bf p}$ is the result of transporting the vector $m{\bf v}$ to the origin $O$ 
(recall we are assuming $m=1$). 

The geometric construction will proceed in two steps.
\begin{enumerate}
\item  
First,  extend the radius vector $OP\equiv {\bf r}$ of $P$ (with the potential
centre $O$ as origin) until it meets the circle $\mathcal D$ at
$I'$. This can be seen as the result of scaling by a factor $({k}/{-E}) ({1}/{r})$, which sends the vector $OP\equiv {\bf r}$ to a new vector with a modulus equal to $({k}/{-E})({1}/{r}) r = ({k}/{-E})$, so this vector tip $I'$ lies on the circle $\mathcal D$. 

\item  
Now consider the image $I$ of $I'$ under reflection with respect to the line $l_P$. 
\end{enumerate}

This construction could be done for
any bounded $E<0$ motion in any bounding arbitrary central potential; as $P$ moves along the orbit, the
point $I'$ moves on $\mathcal D$ and one might expect the point
$I$ to move as well. This is the case for motions in {\itshape other}
central fields, but Kepler motion is {\it exceptional} in this
respect, and we have the following result:

\begin{theorem}
When $P$ moves along a $E<0$  Kepler orbit and
the point $I'$ determined by the previous construction moves on
the circle $\mathcal D$, then the point $I$ stays fixed.
\end{theorem}

In other words, $I$ turns out to be {\itshape independent} of the choice of
the point $P$ on the orbit. 
As we shall see, this geometric `timeless' construction, displayed in Figure \ref{FigvHHelip}, 
will reflect the existence of a constant of motion specific to the Kepler potential.

Before sketching the proof of the Theorem itself, notice that in the figure, the Kepler orbit 
has been already displayed as an ellipse. Actually, the elliptic nature of the orbits
immediately follows as a consequence of the previous theorem:

\begin{corolary}
{\bf (Kepler first law for elliptic orbits)}. The Kepler orbit
with total energy $E<0$ is an ellipse, with a focus at the origin
$O$, the other focus at $I$, and major axis $2a$ equal to the
radius $k/(-E)$ of the circle $\mathcal D$.
\end{corolary}

Proof: $OP$ and $PI'$ are on the same line, hence
$|OP|+|PI'|=k/(-E)$. The reflection in the line $l_P$ is an
Euclidean isometry, so $|PI|=|PI'|$, and then  $|OP|+|PI|=k/(-E)$, so the sum of
distances from $P$ to the fixed points $O$ and $I$ do not change when
$P$ moves on the orbit, and it is equal to the radius of the circle
$\mathcal D$. This agrees with the `gardener' geometric
definition of an ellipse with foci $O$ and $I$.

\medskip

We return to the proof of the theorem which boils down to two
stages: 
\begin{enumerate}
\item   Express the vector ${\bf I}\equiv OI$ in terms of the instantaneous state variables 
of the particle at $P$ 
(i.e.,  position ${\bf r}$ and velocity ${\bf v}$ or momentum ${\bf p}$). 
\item    Compute its time derivative and use the Newton's equations for the Kepler potential 
to check that ${\bf I}$ is indeed a constant of motion. 
\end{enumerate}

As ${\bf I}=OI=OI' - II'$, the first step can
be carried out by evaluating the vectors $OI'$ and $II'$.  $I'$ lies on the circle $\mathcal D$ and then $OI'=\frac{k}{-E}\frac{{\bf r}}{r}$, which immediately leads to 
\begin{equation}  
PI'=OI'-OP=\frac{k}{-E}\frac{{\bf r}}{r} - {\bf r} = \left(\frac{k}{-E}\frac{1}{r} - 1 \right) {\bf r},
\label{PIp}
\end{equation}
and using that  by conservation of energy we have $E=\frac{p^2}{2}-\frac{k}{r}$, we  get
\begin{equation}
PI'=\frac{p^2}{-2 E} {\bf r}. 
\label{PIp2}
\end{equation}

Now, to compute $II'$, we first note that $ {\bf p}\times{\bf L}/p L$ is a unit vector perpendicular to both ${\bf p}$ and ${\bf L}$ (which are also mutually perpendicular), so it has the direction of $II'$. 
The length $MI'$ is the projection of the vector $PI'$ over the line $II'$, and it can be computed as the scalar product $\frac {{\bf p}\times{\bf L}}{p L} \cdot \frac{p^2}{-2 E} {\bf r}$. By using the cyclic symmetry of triple product,  we get ${\bf p}\times{\bf L}\cdot {\bf r} = {\bf r}\times{\bf p}\cdot {\bf L} = {\bf L}^2 = L^2$, and hence we finally have for $II'$ and  {\bf I}
\begin{equation}
  II'=\frac{{\bf p}\times{\bf L}}{-E}=\frac{L}{E} {}^*{\bf p} ,
  \qquad 
    {\bf I} = \frac{k}{-E}\frac{{\bf r}}{r} - \frac{{\bf p}\times{\bf L}}{-E} .
\label{IIf}
\end{equation}

	In order  to check  that ${\bf I}$ is actually a constant of motion we can introduce 
\begin{equation}  
{\bf A} := E \, {\bf I} = {\bf p}\times{\bf L} - k\frac{{\bf r}}{r},
\end{equation}
 and as $E$ itself is a constant of motion,  the second step reduces to checking that ${\bf A}$  is also a constant of motion for the Kepler potential. 
Note that $\dot {\bf L}=0$  and that $\dot{\bf p} = {\bf F}$, then, 
\begin{equation}
  {\frac{d}{dt}{\bf A}}={\bf F} \times {\bf L} - k \frac{d}{dt}{\left(\frac{{\bf r}}{r}\right)}
\end{equation}
with ${\bf F}({\bf r})=-(k/r^2) (\frac{{\bf r}}{r})$ and   a simple direct computation leads to
${\dot{\bf A}}={\bf 0}$.
Of course, ${\bf A}$ is but the standard Laplace-Runge-Lenz vector, the  specific  Kepler constant of motion. 

As stressed by vHH, one merit of this approach is that the specifically Keplerian constant of motion {\it follows directly from the construction}, so the only remaining task is to check it is a constant,  which is the easy part; on the contrary, in the standard approaches, it is not so obvious to figure  out the expression which turn out to be a constant of motion. 

A direct consequence follows from formula (\ref{IIf}): as ${\bf p}$ and ${\bf L}$ are
perpendicular, we have for the modulus of the affine vector  $II'$ the relation $|II'|=\frac{L}{-E}\,p$, (notice that $L$ and $-E$ 
are both positive). This relation, which will be essential for the relation among orbits and hodographs, can be stated as follows: 

\begin{proposition}
For Kepler orbits with $E<0$, as $P$ moves along the orbit, the
Euclidean length $|II'|$ is {\it proportional} to the modulus of
the momentum ${\bf p}$ the particle has when it is at $P$:
\begin{equation}
  |II'| 
  = \frac{L}{-E}\, p
\label{Cor1}
\end{equation}
\end{proposition}

In terms of the geometry of the ellipse, the minor semiaxis length
is   $b = L / (2 \sqrt{-E})$, so the coefficient  $L/(-E)$ in  (\ref{Cor1}) admits an 
alternative expression as $ {L}/({-E}) = {2 b^2}/{L} $.

Then we can sum up these results in two different but equivalent ways: 
\begin{itemize}
\item   In the direct construction, for any point $P$ on the Kepler orbit, produce the radius vector once it 
meets the circle ${\mathcal D}$ at $I'$ and reflect with respect to the tangent line to the orbit at $P$; the reflected point 
$I$ does not depend on $P$. 
 \item   In the reverse construction, choose any point $I'$ on the circle ${\mathcal D}$ and consider the bisector  line of the segment $II'$; this is the tangent to the orbit at some point $P$, and when $I'$ moves along ${\mathcal D}$,   the orbit is recovered as the envelope of  the family of its tangent lines; it is an ellipse with major axis length $2a$, which can   also be described as the set of points equidistant to the fixed point $O$ and the fixed circle ${\mathcal D}$. 
\end{itemize}

We recall that the circle $\mathcal D$, was introduced by vHH as the boundary of the energetically allowed region 
for an orbit with energy $E<0$. 
But now we see from the previous discussion, that the essential property of this circle is precisely 
to be a {\it director} circle ${\mathcal D}_I$  of the ellipse \cite{SomConics24} 
(the director circle companion to the focus $I$).

\subsection{Hyperbolic  orbits}

The geometric approach described in the previous subsection was  only concerned with  elliptic orbits.
The main point was the identification of the circle ${\mathcal D}$ as the director circle of the orbit.
We now extend the approach  to the case $E>0$.

\begin{figure}[h]
\includegraphics[width=\hsize]{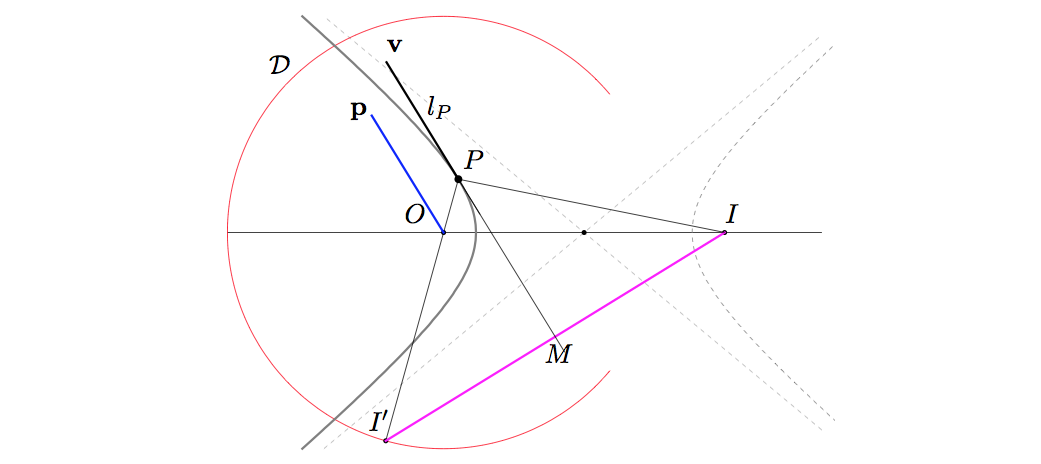}
\vskip-\baselineskip
\caption{\footnotesize \leftskip=20pt \rightskip=20pt 
The extension of the van Haandel-Heckman construction to 
prove that $E>0$ Kepler orbits are hyperbolas. }\label{FigvHHhyperb}
\end{figure}

Now mimic the previous construction for a hyperbolic Kepler orbit ($L\neq 0$ and $E>0$):  

 \begin{enumerate}
\item  
 First, at a point $P$ on the orbit with tangent line $l_P$, {\it scale} the radius vector ${\bf r}$ of $P$ by a factor ${k}/{(-E)}\cdot  {1}/{r}$,  (notice that for $E>0$ this factor is {\itshape negative}). This brings the point $P$  to a new point $I'$ which lies on some circle with center $O$ and radius $k/|E|$, still denoted ${\mathcal D}$  (now $OI'$ has the opposite orientation to $OP$).  In other words, extend the vector $OP$ starting from $O$ {\it in the  opposite sense to ${\bf r}$} until the rescaled vector ${k}/{(-E)} \cdot {{\bf r}}/{r}$ lies precisely on the   circle $\mathcal D$ at a point $I'$ (see Figure \ref{FigvHHhyperb}). 
\item  
  Now consider the image $I$ of $I'$ under reflection with respect to the line $l_P$. 
\end{enumerate}

 Now the main result follows:
 \begin{quote}
{\itshape  As $P$ moves on the orbit, the point $I'$ moves on the circle $\mathcal D$ and the point $I$ stays fixed; from this the hyperbolic nature of the orbit follows. }
\end{quote}
(To be precise, $I'$ moves only on an arc of $\mathcal D$, displayed in continuous red; the remaining, not displayed,  part of the full circle would correspond to the other hyperbola branch, 
which would be the orbit for the repulsive Kepler problem). The reflection
of $I'$ in the tangent line $l_P$ gives a point $I$, which is outside the circle $\mathcal{D}_I$. The result now is that the point $I$  stays at a fixed 
position when $P$ runs the whole Kepler orbit. 

In other words, even in the cases where $E>0$, the orbit is also a conic (here a hyperbola branch) and the circle $\mathcal D$ is a director circle of the conic.  

\subsection{Generic  orbits}

 Now we restate the vHH construction in a way which applies at the same time to  both elliptic and/or hyperbolic orbits. 

Given a Kepler orbit with energy $E\neq0$, for each point $P$ on the orbit, scale the radius vector of the point $P$ with the factor $({k}/{-E}) \cdot ({1}/{r})$, and call $I'$ the point so obtained, which lies on the circle $\mathcal D$ with centre $O$ and radius ${k}/{|E|}$. Now consider the image $I$ of $I'$ under reflection in the line $l_P$ which is the tangent to the orbit at $P$. What singles out the Kepler motion in either the negative or the positive energy regimes is the following result:

\begin{theorem}
When $P$ travels along a Kepler orbit with $E\neq 0$ under the
Kepler central potential $V(r)=-k/r$, and the point $I'$ moves
while lying on the circle ${\mathcal D}$ (whose centre
is $O$ and whose radius is $k/|E|$), then the point $I$ stays at a fixed position.
\end{theorem}

To check that ${\bf I}$ is indeed a constant of motion for any non-zero energy $E$ requires a computation which exactly  mimic the one
performed in the elliptic case. Now for both elliptic and hyperbolic orbits, the relation  between the constant vector ${\bf I}$ which appears naturally in this approach and the standard Laplace-Runge-Lenz vector ${\bf A}$ is $  {\bf I} = {\bf A}/{E}$. 

The reflection with respect to the line $l_P$ is an Euclidean isometry and therefore, $|PI|=|PI'|$, while $PO$ and $PI'$ are
on the same line by construction, but in the present $E>0$ case
there is a slight difference with the previous case: the segment $PO$ is {\it fully
contained} in $PI'$, instead of being two {\it adjacent} disjoint segments,
so that along a positive energy orbit, $|OI'| = |PI'|-|PO| = |PI| -
|PO|$ is a constant length, more precisely equal  to the radius
$2a=k/|E|$ of the circle $\mathcal D$; in this case the quantity which is constant along 
the Kepler orbit is not the sum but the difference of the distances from a generic point on the 
orbit $P$ to the two fixed points $O$ and $I$, and that condition is one of
the classical geometric definitions of a hyperbola. We have got:

\begin{corolary}
{\rm  (Kepler first law for elliptic and hyperbolic orbits)} An $E\neq0$
Kepler orbit  is either an ellipse or a branch of a
hyperbola, with a focus at the origin $O$
and major axis $2a=k/|E|$. The `other' focus $I$ is inside the circle $\mathcal D$ of radius $2a$ for
$E<0$ and outside $\mathcal D$ for $E>0$.
\end{corolary}

\begin{proposition}
As $P$ moves along an  $E\neq0$ Kepler orbit, 
the Euclidean length $|I'I|$ is {\it proportional} to the modulus of the linear momentum 
${\bf p}$ at $P$:
\begin{equation}
  |II'|  = \frac{L}{|E|}\, p =  \frac{2 b^2}{L}\, p.
\end{equation}
\end{proposition}

Here $b$ refers to minor axis length of the conic. 
The proof is identical to that of the case $E<0$
with very minor changes: for instance $OM$ and $IN$ lie on
different sides to the tangent, so here with $\rho_I:=IM,\
\rho_O:=ON$ we have $\rho_I \rho_O=-b^2$ ($b$ is the
hyperbola minor semiaxis length) independently of the choice of the tangent (or
of the point $P$). 

The Laplace-Runge-Lenz vector ${\bf A}$ is a vector at
$O$ which points towards the periastron, with modulus  $A=k\,e$ ($e$ being the eccentricity). This is so for all the signs of the energy (recall $0<e<1$ for negative energy or $e>1$ for positive energy). If now we translate this to the new constant ${\bf I}=\bf A/{E}$, 
we have to discuss the two different generic situations according
as $E<0$ or $E>0$. 

\begin{itemize}
\item     In the $E<0$ case, as $E=-k/(2a)$, the vector ${\bf I}$ points towards the {\it apoastron}, 
and its modulus is $ke/(k/2a)=2ae = 2f$, so this computation confirms the result stated earlier: 
the tip of ${\bf I}$ lies at the ellipse `empty' focus which lie {\itshape inside} $\mathcal D$ as $e<1$. 

\item     In the hyperbolic case, as $E=k/(2a)$, the vector ${\bf I}$ points towards the {\it periastron}, 
and its modulus is again given by $ke/(k/2a)=2ae = 2f$, so as stated before the tip of ${\bf I}$ 
lies at the hyperbola `empty' focus, which, as $e>1$, lies {\itshape outside} the circle $\mathcal D$.  
\end{itemize}

Hence, in all cases, the constant vector ${\bf I}$ points from the origin to the empty focus (and of course, for the parabolic orbits, the modulus of ${\bf I}$ goes to infinity). 
The essential role the circle  $\mathcal D$ plays in this construction is not to be the boundary of the energetically allowed region (which for 
orbits with $E>0$ would be the whole space) but instead to be a director circle ${\mathcal D}_I$ for the conic. We can sum up the results:

\begin{theorem}
{\rm  (Circular character of the Kepler hodograph, \cite{Ham1847})} The hodograph ${\mathcal
H}$
of any Kepler motion is a circle in `momentum space', centred at the point $
{}^*{\!}{\bf A}/L$ and radius $k/L$.
\end{theorem}

We
give a proof within the vHH line of argument. When 
$E\neq 0$, constancy of the vector  
$$ 
  {\bf  I}=\frac{k}{-E}\frac{{\bf r}}{r} - \frac{L {}^*{\bf p}}{-E} =  \frac{1}{E}{\bf A}$$
implies 
$$  
  -{}^*{\bf p}=\frac{\bf  A}{L}+\frac{k}{L}\frac{{\bf r}}{r} \qquad \text{and} \qquad 
  {\bf p}={}^*(-{}^*{\bf p})=\frac{{}^*\bf A}{L}+\frac{k}{L}\frac{{{}^*\bf r}}{r}.
$$
When  $P$ (i.e., ${\bf r}$) moves along the Kepler orbit, this is the
equation of a circle in the ${\bf p}$ space, with centre and radius as stated. 

The `offset' in momentum space between the centre of ${\mathcal
H}$ and the origin point ${\bf p}={\bf 0}$ is ${|{}^*\bf
A|}/{L}=k\,e/L$ and for this reason the vector ${}^*{\!}{\bf A}$ is
called the `eccentricity vector', because the centre is offset
from the origin by a fraction $e$ of the hodograph radius. The linear
momentum space origin $O\equiv {\bf p}={\bf 0}$ is thus inside
${\mathcal H}$ for $E<0$ and outside ${\mathcal H}$ for $E>0$; in
the latter case the actual hodograph is not the complete circle
but only the arc of ${\mathcal H}$ lying in the region ${\bf p}^2> 2
E$: in a hyperbolic motion the modulus of the momentum is always
larger than the modulus of the linear momentum when the particle is `at
infinity'. 

This important result follows from the geometric construction, and the proof underlines the close connection between constancy of ${\bf A}$ and circular character of the hodographs.

The standard proof, dating back to  Hamilton (see e.g. \cite{Mil83}) derives this property from a differential equation obtained from Newton laws by 
changing the time parameter $t$ to the polar angle $\phi$. We have shown that even this step can be dispensed with, as in the vHH approach this circular character of 
hodographs follows from the fact  that  ${\bf I}$ is actually a constant of motion. Actually, this result requires  to use the Newton's equations of motion, so the result does not come from nothing; the point to be stressed is that we must {\itshape use} directly Newton's equations,  but we can completely bypass {\itshape solving} them in any form.

\begin{figure}[h]
\includegraphics[width=\hsize]{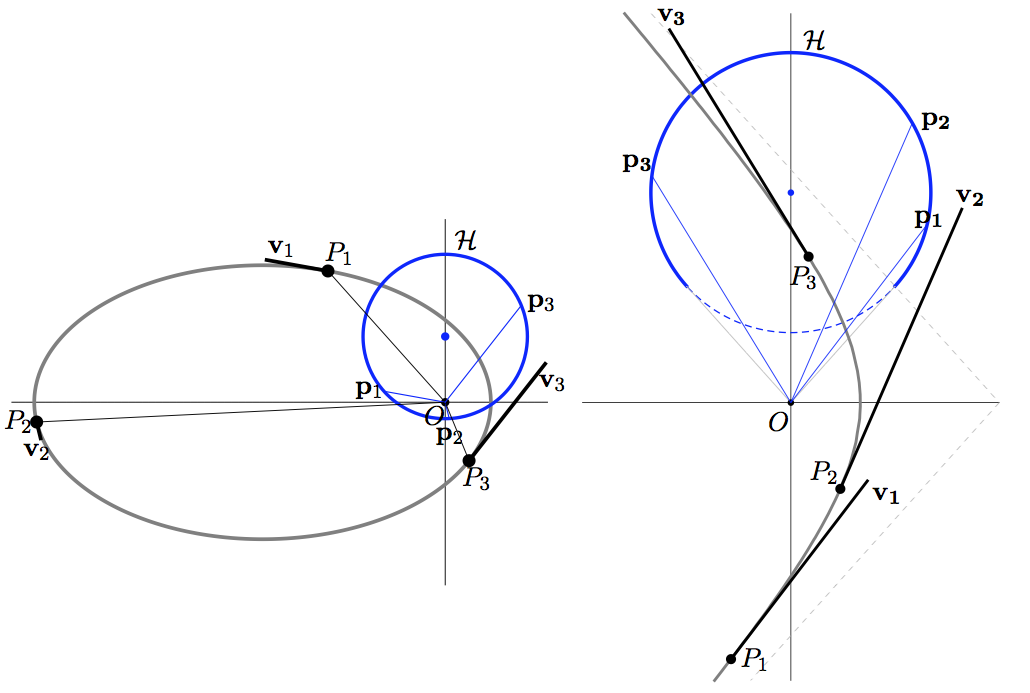}
\vskip-\baselineskip
\caption{\footnotesize \leftskip=20pt \rightskip=20pt 
The hodograph for elliptic $E<0$ and hyperbolic $E>0$ Kepler orbits. }
\label{FigHodographs}
\end{figure}

\subsection{Parabolic  orbits}

The $E=0$ parabolic case may be reached as a limit $E\to 0$ from negative 
or from positive $E$ values.  In both situations, $\mathcal D$ tends to a circle 
with centre at $O$ and infinite radius.  
Thus the original vHH construction degenerates for $E=0$ unless a suitable modification 
is done which allows to deal with this  limit in a regular way. 
One can make a natural choice for this radius so that in the parabolic case 
we get also a working construction (this is described e.g., in the Derbes paper \cite{Derb01}; 
we will not discuss here this question any longer).

As the energy $E$ itself disappears from the hodograph equation (which depends only on ${\bf A}$ and $L$), the result whose proof has been given for $E\neq0$ remains also valid for parabolic orbits. In the $E=0$  parabolic case the hodograph passes through the origin.

\section{Streamlining the geometric construction}

Now we propose a variant of the vHH construction which at the end will simplify it. 
This reformulation turns out to be equivalent to the previous one for
the Euclidean Kepler problem. But this reformulation has some additional interest, 
because it allows a direct extension for the `curved' Kepler problem in a configuration space of constant
curvature, either a sphere or a hyperbolic plane \cite{CRS08JPA, CRS05Kepler, Leonor}.

\begin{figure}[h]
\includegraphics[width=\hsize]{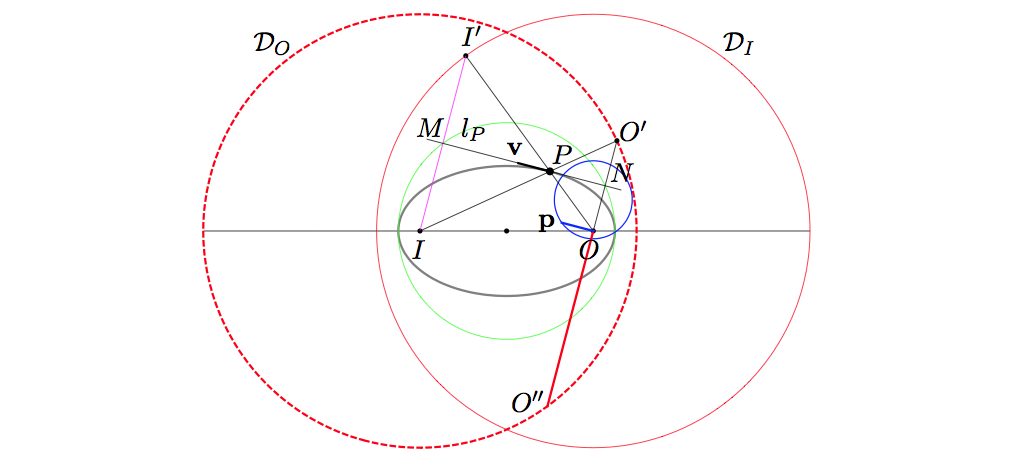}
\caption{\footnotesize \leftskip=20pt \rightskip=20pt 
The `complete' constructions for elliptic $E<0$, Kepler orbits. }\label{FigCompleteEllip}
\end{figure}

\begin{figure}[h]
\includegraphics[width=\hsize]{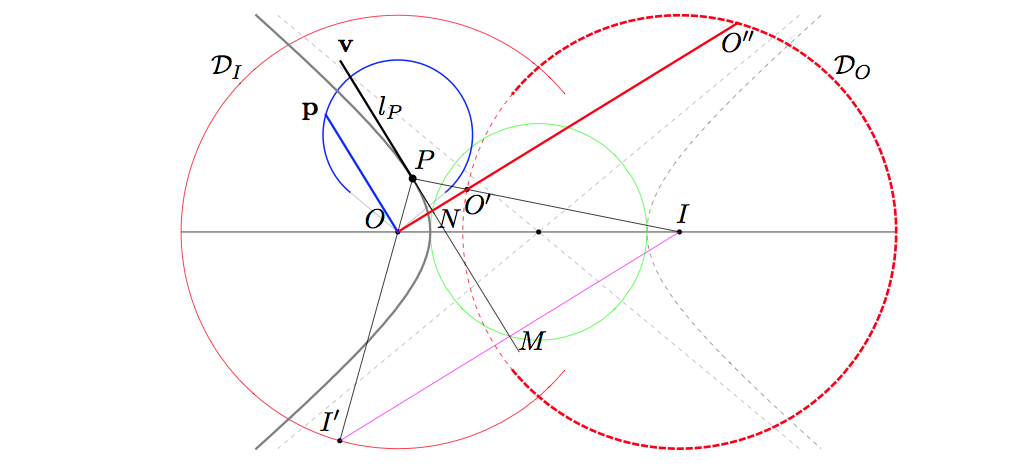}
\caption{\footnotesize \leftskip=20pt \rightskip=20pt
The `complete' constructions for hyperbolic $E>0$ Kepler orbits. }\label{FigCompleteHyperb}
\end{figure}
 
The conics obtained as orbits have not just one pair of matching `director circle -- focus point' but actually two pairs. Further to the director
circle $\mathcal D\equiv{\mathcal D}_I$ associated to the focus
$I$ (which is the only director circle considered up to now), 
there is also another director circle ${\mathcal D}_O$,
`matching' to the focus $O$ and such that the conic
is also the set of points equidistant from ${\mathcal
D}_O$ and $O$. ${\mathcal D}_O$ can be obtained from ${\mathcal
D}_I$ by a central reflection with respect to  the conic centre, and thus
${\mathcal D}_O$ is a circle centred at $I$ and with radius
$2a=k/|E|$. In the figures where both director circles are displayed,  the circle  ${\mathcal D}_O$ is dashed.

Once we know that the generic $E\neq 0$ Kepler orbits are ellipses or
hyperbolas, the  previously described  construction 
can be extended by considering the central reflection with respect to  the centre of the conic. 
This maps the director circle ${\mathcal D}_O$ onto   ${\mathcal D}_I$. 
The image
of $II'$ under this central reflection is $OO''$,
where $O''$ is  the second intersection point of the line $OO'$
with $\mathcal D_O$.
As a consequence of this relation, we may state:
\begin{theorem}
When $P$ travels along a $E\neq 0$ Kepler orbit under the
Kepler central potential $V(r)=-k/r$, then the point $O''$ lies on
the director circle ${\mathcal D}_O$ (whose centre is $I$), and the Euclidean length $|OO''|$ is {\it proportional}
to the modulus $p$ of the linear momentum ${\bf p}$:
\begin{equation}
   |OO''| = \frac{2 b^2}{L} p  = \frac{L}{|E|} p
\end{equation}
\end{theorem}

This result is displayed in Figures \ref{FigCompleteEllip} and \ref{FigCompleteHyperb}, where $II'$ is shown in magenta and its image $OO''$ under the central reflection, in red. Seen as affine vectors,  $II'=-OO''$ and $IO''=-OI'$. The orbit is in dark grey, the director circles in red and dashed red, and the hodograph and the momentum vector are in blue.  In all cases $II'$ and $OO''$ are related by a central reflection with respect to the conic centre and  their equal lengths are proportional to the modulus of the linear  momentum ${\bf p}$. $II'$ is orthogonal to the tangent line $l_P$ while $OO''$ is perpendicular to the linear momentum vector ${\bf p}$.
 
\subsection{Relation with the hodograph}

The next interesting question is to describe the relation between the hodograph and the orbit. Starting from  the circular character of the hodograph, we need a 
construction which applied to the hodograph (whose circular nature is appealing) would allow us to recover the orbit. We mentioned how Feynman did this in a rather descriptive and informal way. 
But now, using the setting provided by the vHH construction, we can describe precisely what Feynman did with full detail through a sequence of three transformations: 

\begin{enumerate}\itemsep=0pt
\item[i)] A rotation by $-\pi/2$ around the origin $O$,
\item[ii)] A homothety around the origin with a scale factor $L/(-E)$ and finally 
\item[iii)] A translation by a vector ${\bf I}$. 
\end{enumerate}

This sequence of transformations can be shown to apply the hodograph ${\mathcal H}$ to the director circle $\mathcal D$ and the linear momentum vector ${\bf p}$ to the vector $II'$. 

We can see that the reformulation of the previous section, which related $E\ne 0$ Kepler motions along the orbit with those of an auxiliary point $O''$ on the director circle ${\mathcal D}_O$, 
allows us to describe this relationship in a simpler way. The important elements in this construction are the rotation by a quarter of a turn, as used by  Feynman \cite{FeynmanLostLecture} 
(but note the opposite sign), and then a homothety; the `translation' step appearing in the Feynman lecture is no longer required, and the two remaining (and now {\it commuting}) steps are enough to relate the
hodograph to the director circle and then to the orbit. 

\begin{theorem}
{\rm  (Relation of Kepler hodograph with the configuration space orbit)} The sequence of the two following transformations
\begin{enumerate}\itemsep=0pt
\item[1)] Rotation by $+\pi/2$ around the origin $O$,
\item[2)] Homotethy around the origin with a scale factor $L/(-E)$, 
\end{enumerate}
applies the hodograph ${\mathcal H}$ to the director circle ${\mathcal D}_O$ and the linear momentum vector ${\bf p}$ on the vector $OO''$. The Kepler orbit corresponding to the hodograph ${\mathcal H}$ is the envelope of  the perpendicular bisectors of the vectors $OO'$ when $O'$ moves along the director circle ${\mathcal D}_O$. Or, alternatively, the Kepler orbit  is the locus of points in configuration space which are equidistant from the origin $O$ and from the director circle ${\mathcal D}_O$. 
\end{theorem}

Before giving the proof, it is worth insisting that the vHH and the Feynman approaches allowed us  to describe the configuration space orbit as the envelope of  a family of lines, 
which were the bisectors of the segments $II'$, as the point $I'$ moves along the director circle $\mathcal D \equiv {\mathcal D}_I$. But the new reformulation, while keeping a similar property 
(the configuration space orbit is the envelope of  the family of the bisectors of the segments $OO'$, as the point $O'$  moves  along the director circle ${\mathcal D}_O$)  
allows us a more direct description of the configuration space orbit: it is the set of points in configuration space which are equidistant from the fixed point $O$ (the centre of forces) and from the fixed circle ${\mathcal D}_O$.

\begin{figure}[h]
\includegraphics[width=\hsize]{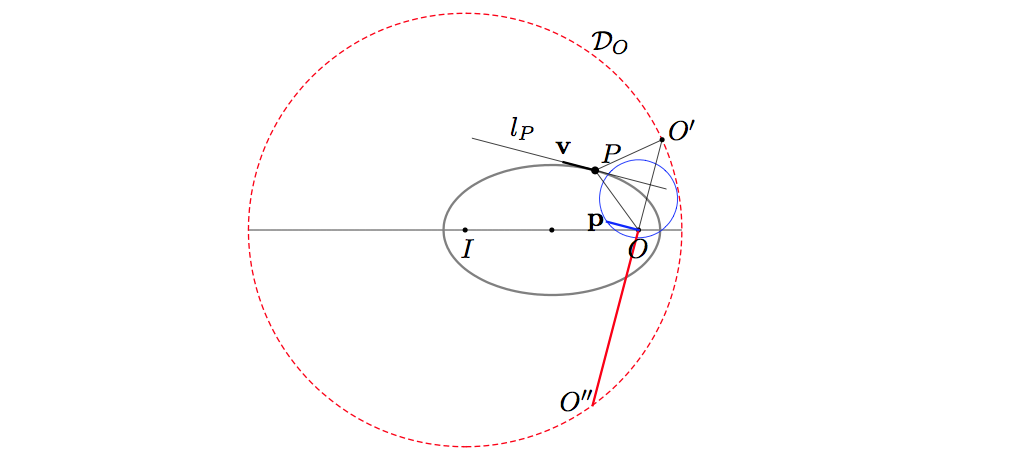}
\vskip-\baselineskip
\caption{\footnotesize \leftskip=20pt \rightskip=20pt
The `minimal' constructions for
elliptic $E<0$ Kepler
orbits. }\label{FigMinEllip}
\end{figure}

\begin{figure}[h]
\includegraphics[width=\hsize]{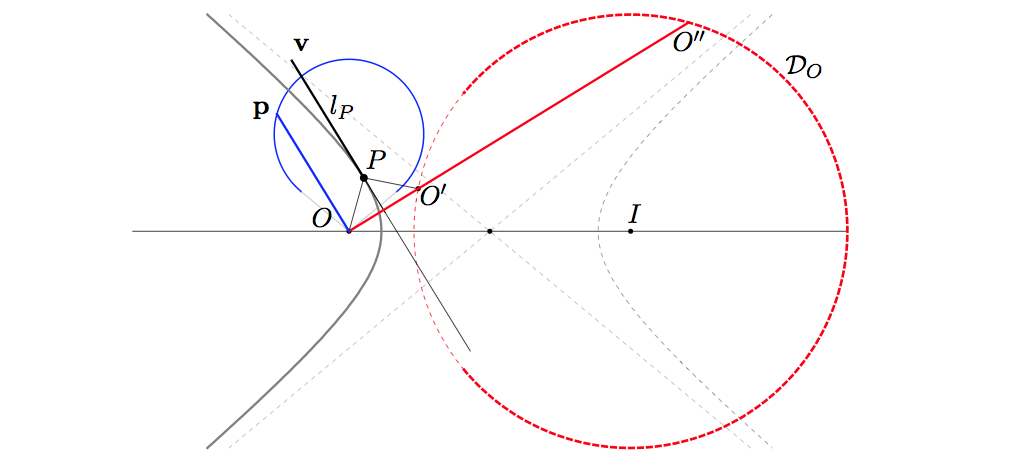}
\vskip-\baselineskip
\caption{\footnotesize \leftskip=20pt \rightskip=20pt
The `minimal' constructions for hyperbolic $E>0$ Kepler orbits. }
\label{FigMinHyperb}
\end{figure}

The proposition follows by direct computation: for any vector ${\bf w}$ in momentum plane, the two steps make the transformations:
\begin{equation}
  {\bf w} \longmapsto  {}^*{\bf w} \longmapsto \frac{L}{-E}\,{}^*{\bf w}.
\end{equation}
Under the composition of the two steps, a generic point on the hodograph ${\bf p}$ goes to
\begin{equation}
  {\bf p} \longmapsto  {}^*{\bf p} \longmapsto   \frac{L}{-E}\,{}^*{\bf p}
   =  \frac{\bf A}{E} - \frac{k}{|E|}\frac{\bf r}{r}  \equiv  OO'.
\end{equation}
Notice that $OO''$ is automatically perpendicular to ${\bf p}$;
the hodograph centre ${}^*{\!}{\bf A}/L$ goes to
\begin{equation}
  \frac{1}{L} \,{}^*{\!\bf A} \longmapsto  -\frac{1}{L}{\bf A} 
  \longmapsto  \frac{1}{E}{\bf A},
\end{equation}
which means that under the two steps 
the hodograph becomes the
director circle ${\mathcal D}_O$, with radius ${k}/{|E|}$ and
centre at $I$. The origin of the linear momentum space, ${\bf p}={\bf 0}$,
stays fixed.

This is depicted in Figures  \ref{FigMinEllip} and \ref{FigMinHyperb}, where the director
circle ${\mathcal D}_I$ and all their associate elements have been
removed because they are not actually relevant for this streamlined construction. The orbit itself is in dark grey, the director circle $\mathcal{D}_O$   in dashed red, and the hodograph and the linear momentum vector are
in blue. The relation between the director circle ${\mathcal D}_O$
and the hodograph, by a rotation of $\pi/2$ and a scaling with
factor $L/(-E)$ is clearly displayed; the sign of this scaling factor depends on the sign of the energy.

This sequence of transformations can be shown to apply the hodograph ${\mathcal H}$ to the director circle $\mathcal D$ and the momentum  ${\bf p}$ to the vector $II'$. We now see that the reformulation of the previous section, which related $E\ne 0$ Kepler motions along the orbit with those of  auxiliary points $O''$ on the director circle ${\mathcal D}_O$,  allows us to describe this relationship in an even simpler way. The `translation' step of Feynman lecture is no longer required, and the two remaining (and now {\it commuting})  steps are enough to relate the hodograph to the director circle and then to the orbit. The relevant elements in this construction are the rotation by a quarter of a turn, as used by  Feynman \cite{FeynmanLostLecture}, and then a homothety.

The translation iii) in the Feynman
relation among hodograph and orbit only serves to map the director
circle ${\mathcal D}_O$ onto ${\mathcal D}_I$, and thus it is
unnecessary. The `correct' relation among both director circles,
swaping ${\mathcal D}_O$ for ${\mathcal D}_I$ and $OO''$ for $II'$
is not actually a translation, but a central reflection with respect to the
ellipse or hyperbola centre. As this can be suitably decomposed as a
product of a central reflection with respect to $O$ and a translation with
vector $OI$, this is the reason for the 
opposite signs at stage 1) of Theorem 5, as compared with the sign of the rotation angle $-\pi/2$ 
in the stage i) of the Feynman lecture.

\section{A  comment on the Kepler problem on curved spaces}

The idea that the Kepler problem  (and also the harmonic oscillator) can be correctly defined on constant curvature spaces appears in a  book of Riemannian geometry of 1905 by  Liebmann \cite{Liebmann}; 
but it was Higgs \cite{Hig79}  who studied this system with detail in 1979 (the study of Higgs was limited 
to a spherical geometry but his approach can be extended, introducing the appropriate changes,  
to the hyperbolic space).  Since then several authors have studied the Kepler problem on curved spaces and have analysed the 
existence of dynamical symmetries leading to constants of motion that can be considered as the appropriate 
generalisations of the Euclidean Laplace-Runge-Lenz vector.    In addition, it has also been proved, by introducing  a modified version of the change $u=1/r$, 
the existence of a curved version of the well known  Binet equation (see \cite{CRS05Kepler} and references therein). 

At this point it seems natural to ask whether the Kepler problem on  curved spaces 
(of a constant curvature) can be analysed by the use of  an  approach similar to the one 
presented in previous sections (that is, without integral calculus or differential equations). 

At first sight, the answer seems to be negative. The hodograph 
(defined starting from the velocity vectors) seems to involve an implicit 
transport of the velocity vector at each point of the orbit to a 
common origin $O$. In a flat configuration space, parallel transport 
is uniquely defined, no matter of which path is followed, and this 
makes irrelevant the question about `where are these vectors applied', either 
at each point on the orbit $P$ or in a common origin $O$. 
Thus, to try to extend a `velocity based hodograph approach' to a constant curvature configuration 
space might seem pointless,  because the result of this transport would depend on the path 
followed, and hence this 'velocity based hodograph' itself seems to be 
not well defined. This is of course true.

But the point to be stressed is that the true hodograph 
should be based in the momenta ${\bf p}$ rather than in the velocity. Indeed the 
paralell transport is an inessential element in the construction 
which is only required if one starts with the velocity and not 
with the Noether momenta ${\bf P}$, as one should.  In the 
construction presented in the previous paragraphs, the most 
important vector is ${\bf p}$, which is a vector at $O$ (see  Figures \ref{FigvHHelip}, \ref{FigvHHhyperb}, \ref{FigHodographs}) and coincides with the Noether moment. As the 
Euclidean parallel transport is path independent, this vector 
at $O$ {\itshape coincides} with the parallel transported of 
$v$ along {\itshape any path joining $P$ to $O$}. But in a space 
of a constant curvature, while the result of some unqualified parallel transport 
of the velocity vector to $O$ would be undefined, the components of Noether 
momenta ${\bf P}$ are still well defined, and they are, alike the components of the the other 
(conserved) Noether momentum ${\bf L}$, a vector at $O$. Nevertheless 
if in a constant curvature space everything is written in terms 
of the associated momenta (which are naturally vectors in an auxiliary
space), it turns out that  both Theorems 4 and 5 have a direct 
extension to this case. 

Henceforth, the construction we have here described 
allows a quite direct extension to the case of constant curvature 
configuration space.  This will be discussed elsewhere. 

\section{Final comments }

The Kepler problem is studied in all  books of Classical Mechanics and it is solved by
making use of integral calculus and differential equations (e.g., Binet equation). 
Nevertheless the Newton approach presented in the {\it Principia} was mainly related 
with the classical language of Euclidean geometry. 
This property (that it can be solved by the use of a purely geometric approach)  
is a specific property of the Kepler problem that distinguish it from all the other problems 
with central forces. 
This simplicity is a consequence of the existence of 
an additional constant of motion which is specifically Keplerian: the Laplace-Runge-Lenz vector. 
In fact, the circular character of the Kepler hodograph, discovered and studied by Hamilton, 
is just a consequence of the existence of this additional integral of motion.

In the first part of this paper we have   reviewed and compared two geometric approaches 
to the Kepler problem,  which were originally devised for only dealing with   elliptic orbits. 
They are due to Feynman and to van Haandel-Eckman. 
Both fall into the broad class of `hodograph approaches'  but the vHH one somehow reverses 
the usual logic in a way which avoids the recourse to any differential equation, so making this 
approach accessible to a wider audience. 
In particular, the vHH approach leads in a natural and purely algebraic way to the specifically 
Keplerian constant of motion,  the Laplace-Runge-Lenz vector. 

Then taking this as starting point, we identify the important geometric role 
of some circles (director circles) entering into these constructions. 
First, we show that both approaches can be suitably extended to cover, 
not only bounded elliptic orbits,  but also  open hyperbolic ones. 
And second, by making use of the properties of these director circles, the full analysis is streamlined, 
so that the final 'minimal' description of the relationship of the hodograph with the true orbit 
in configuration space is neater than in the previous ones. 

The conic nature of the orbit follows from this approach in a purely algebraic way, 
and this applies both to elliptic and hyperbolic orbits. In summary, 
this can be very suitable for beginning students, as the Newton laws are simply used directly, 
but no explicit solving of any differential equation is required.

\section*{Acknowledgements}

JFC and MFR acknowledge support from research projects MTM-2012-33575 (MEC, Madrid) and 
DGA E24/1 (DGA, Zaragoza) and MS from research project MTM2014-57129 (MINECO, Madrid).


{\small

 }
\end{document}